# The electron-phonon relaxation time in thin superconducting titanium nitride films.

A. Kardakova[1,5,a)], M. Finkel[1], D. Morozov[2], V. Kovalyuk[1], P. An[1], C. Dunscombe[2], M. Tarkhov[3], P. Mauskopf[2], T.M. Klapwijk[1,4], and G. Goltsman[1,6]

[1]Physics Department, Moscow State Pedagogical University, Moscow, 119991, Russia

[2]School of Physics and Astronomy, Cardiff University, Cardiff, CF24 3AA, UK

[3]National Research Centre "Kurchatov Institute", Moscow, 123128, Russia

[4]Kavli Institute of Nanoscience, Delft University of Technology, 2628 CJ, Delft, Netherlands

[5]Moscow Institute of Physics and Technology (State University), Dolgoprudny, 141700, Russia

[6]National Research University Higher School of Economics, Moscow, 101000, Russia

a) Author to whom correspondence should be addressed. Electronic mail: kardakova@rplab.ru.

ABSTRACT:

We report on the direct measurement of the electron-phonon relaxation time, $\tau_{eph}$, in disordered TiN films. Measured values of $\tau_{eph}$ are from 5.5 ns to 88 ns in the 4.2 to 1.7 K temperature range and consistent with a $T^{-3}$ temperature dependence. The electronic density of states at the Fermi level $N_0$ is estimated from measured material parameters. The presented results confirm that thin TiN films are promising candidate-materials for ultrasensitive superconducting detectors.

For several decades titanium nitride (TiN) has been the subject of active research. It has a number of attractive properties such as ultra-high hardness, high wear- and corrosion-resistance, mechanical robustness and high thermal conductivity. All of these make them also suitable for use in nanoscale structures. Bulk TiN material is a superconductor with a transition temperature below 6 K[1]. Experiments have demonstrated that the critical temperature of thin TiN films decreases with decreasing thickness, i.e. with increase of the sheet resistance (in the normal state)[2,3]. Furthermore the critical temperature of the TiN films can be controlled over a wide range (0<$T_c$<5 K) by varying the $N_2$ concentration during the deposition[1,4,5]. The superconducting transitions remain sharp over the whole range of $T_c$ values. Recent studies of low temperature transport properties of TiN superconducting films thinner than 5 nm show a disorder–driven transition from a superconductor to an insulating phase[6]. It is an interesting physics problem, which may be used for sensors and detectors. The strongly disordered TiN films are also ideal materials for the observation of quantum coherent phase slips[7]. Moreover due to the attractive superconducting properties and high resistivity enabling efficient photon absorption the thin superconducting TiN films have been successfully utilized in the development of ultrasensitive THz detectors, such as hot-electron microbolometers[8], microwave kinetic inductance detectors (MKIDs) with coplanar waveguide resonators[9] and lumped-element resonators[10]. This type of detectors achieves optical noise equivalent power (NEP) of $3.8 \times 10^{-19}$ W/Hz$^{1/2}$ allowing background limited sensitivity for astronomical observations in the THz wavebands[11,12]. Nowadays the MKID is the most promising superconducting detector for astronomical instruments due to the scalability and the possibility of multiplexing a large number of pixels. TiN films are very attractive for these MKIDs[10]. In all these applications, the essential characteristics of the devices such as noise in detectors and decoherence time in qubits are strongly dependent on the energy relaxation process in the material. An accurate knowledge of the relaxation rate is needed for the successful development of the devices.

The relaxation process in the superconducting state is mainly governed by two aspects. First, the quasiparticles need to emit or absorb energy in excess of the gap 2Δ to be recombined or excited, involving the interaction with phonons. Secondly, the number of quasiparticles has to be very small well below the critical temperature. This all leads to exponentially slow quasiparticle recombination rates at low temperatures as described by the general expression[13]

$$\tau_{rec}^{-1} \propto \tau_{eph}^{-1}(T_c) \sqrt{\frac{T}{T_c}} e^{-\frac{\Delta}{k_B T_c}}. \qquad (1)$$

Here $\tau_{rec}$ is the recombination time, $\tau_{eph}$ is the electron-phonon interaction time at Tc, $k_B$ is Boltzmann's constant, T is the bath temperature, $\Delta$ is the superconducting gap. Indeed, the experimental dependencies of $\tau_{rec}(T)$ are described by this expression at the temperatures down to $T/T_c \approx 0.175$[14], after which some, currently not fully understood, saturation sets in. In the other limit, when the temperature approaches $T_c$ for sufficiently thin films the relaxation process is usually controlled by the material dependent electron-phonon interaction (e-ph), which means that $\tau_{rec}$ at $T_c$ is approximately equal to $\tau_{eph}$. The determination of the e-ph time in TiN is our main focus here.

In this letter we report direct measurements of the e-ph time in thin disordered TiN films in the 1.7-4.2 K temperature range, which is as well a typical range for the needed $T_c$'s for MKIDs. Also we estimate the electron density of states $N_0$ of the films. The results confirm that the thin superconducting TiN is a promising material for the different detector technologies.

We study thin TiN films, which were deposited on a sapphire substrate, held at ambient-temperature, by dc magnetron sputtering from a pure Ti target with a power of 900 W in an Ar-$N_2$ plasma. The background pressure of the system was $2 \times 10^{-7}$ mbar. The Ar flow rate and the $N_2$ flow rate were set at 1.3 sccm and 6.5 sccm, respectively. High resolution Auger-electron spectroscopy (the PHI 700 Scanning Auger Nanoprobe) revealed that the films are homogeneous in thickness and the atomic ratio between titanium and nitrogen is 0.43/0.57, respectively, and the concentration of oxygen is about 10%. The films have thicknesses of 80 nm, 22 nm and 15 nm with the critical temperatures of 4.6 K, 3.6 K and 2.6 K, respectively, with $\Delta T_c$=0.1 K ( we refer to them as N1, N2 and N3). The resistivity of the films was derived from the sheet resistance R□ at 300 K and film thickness d, as $\rho = R_\square d$. We also measured the electron diffusion constants for these films. The value of D was experimentally determined from the temperature dependence of the second critical magnetic field $H_{c2}$ using[15]:

$$D = -\frac{0.407 \pi k_B}{e}\left(\frac{\partial H_{c2}}{\partial T}\right)^{-1}\bigg|_{T=T_c}. \qquad (2)$$

The parameters of the films are summarized in Table 1.

The films were patterned into strips of 2 μm wide and 14 μm long with large TiN contacts, using photolithography and chemical plasma etching with $SF_6$. The RF power of the etching process was 50 W with working pressure of 5mTorr, the Ar and the $SF_6$ flow rates were set to 25 sccm and 15 sccm, correspondingly.

The energy relaxation time can be directly determined from a hot-electron measurements. The schematic block diagram of the experimental setup is shown in Figure 1. In the experiment the sample is held at a temperature in the middle of the superconducting transition, and biased with a small dc current. The sample is exposed to the amplitude modulated submillimeter electromagnetic (EM) radiation with the modulation frequency $\omega_m$. The absorbed radiation power causes an increase of the electron temperature $T_e$, which leads to an increase of the resistance of the sample as well, producing a voltage signal proportional to the bias current $\delta V = I \times \delta R$. We use the dependence of output voltage $\delta V$ on $\omega_m$ to determine the energy relaxation time. As EM radiation sources we used two backward wave oscillators (BWO) with carrier frequencies $f_{LO}$=300 GHz and 140 GHz that satisfy, as will be shown, the condition $\omega_{LO} \gg 1/\tau_{eph}$. The amplitude modulation at frequencies 0.1-10 MHz was achieved by the modulation of the BWO power with the anode voltage $V_a$ of the BWO. The value of $V_a$ is selected to correspond to the maximum derivative of dP/dV$_a$ of the BWO operation-characteristic. When an alternating voltage with the amplitude of several volts is superimposed on the anode voltage, the amplitude modulation with the frequency of the alternating voltage is obtained. The amplitude modulation at frequencies larger than 5 MHz was obtained by the beats of the oscillation of two BWOs operated at nearby frequencies. The output voltage $\delta V(\omega_m)$ and the frequency $\omega_m$ were measured using a spectrum analyzer. To avoid excessive heating of the samples the total incident power from DC bias and BWOs was kept small at the level where its change has negligible effect on output signal. To investigate the temperature dependence of $\tau$ we used the magnetic field H that allows to shift the superconducting transition to the temperatures below the Tc at H=0. Further details of the method are available in Refs.16 and 17.

The typical measured dependencies $\delta V(\omega_m)$ are presented on Figure 2. We applied a least square method fit to the measured dependencies according to the equation

$$\delta V(\omega_m) = \delta V(0)/\sqrt{1 + \omega_m^2/\omega_{3dB}^2}, \qquad (3)$$

where $\omega_{3dB} = 1/\tau$ is the 3dB roll-off frequency of the signal, and $\tau$ the energy relaxation time which we identify as the electron-phonon time.

The identification of the roll-off frequency and the energy relaxation time is applicable when the electron-electron scattering is faster than the electron-phonon interaction. As illustrated in the inset of Figure 2 the electron-subsystem can be described by the Fermi distribution function with an effective electron temperature $T_e$ exceeding the thermostat temperature $T_{bath}$. Also, the phonons in the film can be considered as a thermostat on the condition $\frac{c_e}{c_{ph}}\tau_{esc} \ll \tau_{eph}$, where $c_e$ and $c_{ph}$ are the electron and phonon specific heat capacities, respectively, and $\tau_{esc}$ is the time which corresponds to the Kapitza resistance. If the excitation power is low enough that $T_{ph} - T_{bath} \ll T_e - T_{bath} \ll T_{bath}$, where $T_{ph}$ is the phonon temperature, the equation describing the phonons temperature dynamics is dropped. In this case the energy dynamics of the electron subsystem exposed to electromagnetic radiation is described with one linearized heat flow balance equation:

$$c_e \Omega \frac{dT_e}{dt} = -\frac{c_e}{\tau_{eph}}\Omega(T_e - T_{bath}) + I^2 \frac{\partial R}{\partial T_e}(T_e - T_{bath}) + I^2 R + P_{RF}. \qquad (4)$$

The left side describes change in the electron subsystem energy, the first term on the right side represents the heat flow from electrons to the bath, the second term describes the electro-thermal feedback, $I^2 R$ is the Joule power, $P_{RF} = P_0(1 + \cos\omega_m t)$ is the absorbed power of radiation and $\Omega$ is volume. According to the Eq.4 the amplitude $\delta T_e$ of the electron temperature alternating at the modulation frequency is given by $\delta T_e = \frac{P_0}{c_e}[1 + (\tau\omega_m)^2]^{-\frac{1}{2}}$, where $\tau = \tau_{eph}(1-\alpha)^{-1}$ when the effect of self-heating by the transport current can be neglected according to the condition $\alpha = \frac{I^2}{c_e}\frac{\partial R}{\partial T_e}\tau_{eph} \ll 1$. The value of a bias current was chosen to satisfy the condition α<0.2, so that the value of the measured τ did not change with a further lowering of the bias current. It should be noted that the dependencies $\delta V(\omega_m)$ of the device response are well fitted with the single roll-off Eq.3. Hence, it presumably indicates that the $\tau_{esc}$ is not appearing for the reported film thicknesses and $\tau_{eph}$ is the bottleneck in the energy relaxation process.

The electron energy relaxation times for samples N1, N2 and N3 obtained at different temperatures are shown in Figure 3. Data sets are fitted with the power law dependence $\tau \sim T^{-n}$ with $n = 3.05 \pm 0.14$ for N1 and $n = 2.84 \pm 0.38$ for N3. The best fit for all data on the Figure 3 gives the time values as 5.5 ns at T=4.2 K and 88 ns at T=1.7 K and corresponds to the temperature dependence $\tau_{eph} = \alpha T^{-3.0 \pm 0.13}$, with $\alpha = 407\ ns \times K^3$ as a material parameter.

Since the time constant in all samples depends on the temperature and is found to be independent of the film thickness, it strongly suggests that the energy relaxation rate is purely limited by the electron-phonon interaction time. The diffusion of hot electrons to the large contacts is not taken into account here, as the electron diffusion length $L_e = \sqrt{\tau D}$ is considerably smaller than the length of the samples.

The same power n=3 for TiN films has been derived for transition-edge devices (TES) in the temperature range 0.015÷0.05 K[8]. A time constant of 5 ms has also been directly measured for TiN at 50 mK by adding a small pulse onto the voltage bias of TES microbolometers. This agrees with the value obtained from our experimental data by extrapolating to 50 mK. For comparison, we list the electron-phonon times for some other materials suitable for the superconducting detectors. The $\tau_{eph}$ ranges from 1 ns to 10 ns at 4.2 -1.7 K in Nb[16] with a temperature dependence with the power n=2. For Ti[18] and Hf[18] $\tau_{eph}$ is reported in the range from 1 μs to 30 ms at 0.5 K÷0.03 K with n=4 and $\tau_{eph}$ is changing from 10 ps to 200 ps at 10.5 – 1.7 K with n=1.6 in NbN[19].

The power n=3 is predicted by the theory for clean metals when reduced dimensionality effects are not important[20], but for films with strong elastic scattering the power is expected[21,22] to be n=4. The power value of n=4 is also observed experimentally[18,23] but usually the values of n differs from the theory, which is attributed to a not fully achieved dirty limit. In our case, the dirty limit in the sense of $q_T l \ll 1$ (where $q_T = k_b T / \hbar u$ is the wave number of thermal phonons, u is the sound velocity, ℓ is the electron mean free path) is achieved already at T≈ 4.2 K with $q_T \ell$=0.1. The values of the mean free path are estimated with $l = 3D/v_F$ with the Fermi velocity $v_F = 7 \times 10^5$ m/s for TiN films[24] and are listed in Table I. The fact that we find n=3 rather than n=4 may be due to the fact that the predicted temperature dependence is based on the Debye phonon spectrum[21,22]. The real phonon spectrum in our TiN films is unknown, since it may also be modified by strong disorder as well as due to acoustic interaction of the film with the substrate.

In addition we also list the electron density of states at the Fermi level, $N_0$. $N_0$ is another important material parameter that determines the design and, consequently, the noise performance of MKIDs and hot electron bolometers (HEBs)[25]. The electron densities of states for the three samples are estimated from experimental values of the resistivity and the electron diffusion constant using expression $N_0 = 1/(e^2 \rho D)$ and are shown in Table I. These values of $N_0$ are greater than the values previously estimated[10,26]. The value of $N_0$ is a useful parameter for calculating an energy resolution of detectors $\delta \varepsilon = NEP \sqrt{\tau}$. The corresponding energy resolution of HEB and TES can be found as $\delta \varepsilon \approx \sqrt{k_B T^2 c_e}$, where $c_e = \gamma V T$ is the electron heat capacity with $\gamma = \frac{\pi^2}{3} k_B^2 N_0$. For MKIDs it can be written[27] as $\delta \varepsilon \sim \sqrt{k_B \Delta^2 N_0 V/(\alpha_{sc} Q_{i,max})}$, where V is the active volume of the detector, $\alpha_{sc}$ is the kinetic inductance fraction and $Q_{i,max}$ is the maximum quality factor of the resonator. For typical values of V≈0.01 μm³ for HEBs[25] the energy resolution δε is about of $10^{-3}$ aJ at 0.3 K and for MKIDs[28] with the values of V≈70 μm$^3$, $α_{sc}$≈1, $Q_{i,max}$≈10$^5$, Tc=1 K the value of δε is order of 1.5×10$^{-3}$ aJ at 0.1 K. These estimates for TiN films indicate that photon counting in the far-infrared and THz should be feasible for detectors at low temperatures.

In conclusion, we studied the electron-phonon interaction time in sputtered thin TiN films. We find that the temperature dependence of $τ_{eph}$ for three films with different thicknesses corresponds to the same trend of T$^{-3}$. The experimentally determined parameters of the electronic subsystem of the TiN films confirm that they are ideal for ultrasensitive superconducting detectors.

ACKNOLEDGEMENTS

The work was supported by the Ministry of Education and Science of the Russian Federation, contract 14.B25.31.0007 and by the RFBR grant 13-02-91159.

REFERENCES

[1] W. Spengler, R. Kaiser, A.N. Christensen, and G. Muller-Vogt, Phys. Rev. B **17**, 1095 (1978).

[2] W. Tsai, M. Delfino, J.A. Fair, and D. Hodul, J. Appl. Phys. **73**, 4462 (1993).

[3] B. Sacepe, C. Chapelier, T.I. Baturina, V.M. Vinokur, M.R. Baklanov, and M. Sanquer, Phys. Rev. Lett. **101**, 157006 (2008).

[4] L.E. Toth, C.P. Wang, and C.M. Yen, Acta. Metall. **14**, 1403 (1996).

[5] T.P. Thorpe, S.B. Quardi, S.A. Wolf, and J.H. Claassen, Appl. Phys. Lett. **49**, 1239 (1986).

[6] T.I. Baturina, Yu. Mironov, V.M. Vinokur, M.R. Baklanov, and C. Strunk, Phys. Rev. Lett. **99**, 257003 (2007).

[7] O.V. Astafiev, L.B. Ioffe, S. Kafanov, Yu. A. Pashkin, K. Yu. Arutyunov, D. Shahar, O. Cohen, J.S. Tsai, Nature **484**, 355 (2012).

[8] P. Day, H.G. Leduc, C.D. Dowell, R.A. Lee, A. Turner, and J. Zmuidzinas, Journal of Low Temperature Physics **151**, 477 (2008).

[9] M.R. Vissers, J. Gao, D.S. Wisbey, D.A. Hite, C.C. Tsuei, A.D. Corcoles, M. Steffen, and D.P. Pappas, Appl. Phys. Lett. **97**, 232509 (2010).

[10] H.G.Leduc, B. Bumble, P.K. Day, B.Ho Eom, J. Gao, S. Golwala, B.A. Mazin, S. McHugh, A. Merrill, D.C. Moore, O. Noroozian, A.D. Turner, and J. Zmuidzinas, Appl. Phys. Lett. **97**, 102509 (2010).

[11] R.M.J. Janssen, J.J.A. Baselmans, A. Endo, L. Ferrari, S.J.C. Yates, A.M. Baryshev, T.M. Klapwijk, Appl. Phys. Lett. **103**, 203503 (2013).

[12]P.J. de Visser, J.J.A. Baselmans, J. Bueno, N. Llombart, T.M. Klapwijk, arXiv:1306.4238, accepted for publication in Nature Communications.

[13]S.B. Kaplan, C.C. Chi, D.N. Langenberg, J.J. Chang, S. Jafarey, D.J. Scalapino, Phys. Rev. B **14**, 4854 (1976).

[14]R. Barends, J.J.A. Baselmans, S.J.C. Yates, J.R. Gao, J.N. Hovenier, and T.M. Klapwijk, Phys. Rev. Lett. **100**, 257002 (2008).

[15]Barry P. Martins, *New frontiers in Superconductivity Research* (Nova Science Publishers, New York, 2006), p.171.

[16]E.M. Gershenzon, M.E. Gershenzon, G.N. Gol'tsman, A.M. Lyul'kin, A.D. Semenov and A.V. Sergeev, Zh. Eksp. Teor. Fiz. **97**, 901 (1990).

[17]E.M. Gershenzon, M.E. Gershenzon, G.N. Gol'tsman, A.D. Semenov and A.V. Sergeev, Zh. Eksp. Teor. Fiz. **86**, 758 (1984).

[18]M.E. Gershenson, D. Gong, T. Sato, B.S. Karasik and A.V. Sergeev, Appl. Phys. Lett. 79, 2049 (2001).

[19]Yu. P. Gousev, G. N. Gol'tsman, A. D. Semenov, E. M. Gershenzon, R. S. Nebosis, M. A. Heusinger, and K. F. Renk, J. Appl. Phys. **75**, 3695 (1994).

[20]S.-X. Qu, A.N. Cleland, M.R. Geller, Phys. Rev. B. **72**, 224301 (2005).

[21]A. Schmid, in *Localization, Interaction, and Transport Phenomena 1985: Proceedings of the International Conference,* Braunschweig, Fed. Rep. of Germany, 23 August – 28 August 1984, edited by B. Kramer, G. Bergmann, Yv. Bruynseraede (Springer Berlin Heidelberg, Berlin, 1985) **61**, pp. 212-22.

[22]M. Reizer and A. Sergeev, Zh. Eksp. Teor. Phys. **90**, 1056 (1986); A. Sergeev, V. Mitin, Phys. Rev. B **61**, 6041 (2000).

[23]L.J. Taskinen and I. J. Maasilta, Appl. Phys. Lett. 89, 143511 (2006).

[24]J. S. Chawla, X. Y. Zhang and D. Gall, J. Appl. Phys. 113, 063704 (2013).

[25]J. Wei, D. Olaya, B.S. Karasik, S.V. Pereverzev, A.V. Sergeev, M.E. Gershenson, Nature Nanotechnology **3**, 496 (2008).

[26]V. Ern, A.C. Switendick, Phys. Rev. B **137**, 1927 (1965); R. Ahuja, O. Eriksson, J. M. Wills, B. Johansson, Phys. Rev. B **53**, 3072 (1996).

[27]The estimation for MKID energy resolution is obtained with using the expression[10]

$$NEP = 2\eta_{opt}\sqrt{2\eta_{read}N_0\Delta^2 V k_B T_{amp}/(\chi_c \alpha_{sc} S_1 \tau Q_{i,max})},$$ where the efficiency of photon energy to quasiparticle conversion $\eta_{opt} \approx 0.7$, the efficiency of read-out power to quasiparticle conversion $\eta_{read} \leq 1$, the parameter of matching the coupling and internal

quality factors $\chi_c \leq 1$, the amplifier noise temperature $T_{amp} \approx 1 - 10\ K$, the Mattis-Bardeen factor $S_1 \approx 1$.

[28]J.Gao, M.R. Vissers, M.O. Sandberg, F.C.S. da Silva, S.W. Nam, D.P. Pappas, K.D. Irwin, D.S. Wisbey, E. Langman, S.R. Meeker, B.A. Mazin, H.G. Leduc, and J. Zmuidzinas, Appl. Phys. Lett. **101**, 142602 (2012).

TABLE I. Parameters of the film samples.

| | Thickness, nm | $T_C$, K | $\Delta T_C$, K | $R_\square^{300K}$, Ohm/sq | $\rho^{300K}$, μOhm cm | $j_c$ (T=1.7 K), A/cm$^2$ | $D$, cm$^2$/s | $N_0$, eV$^{-1}$ μm$^{-3}$ | $l$, nm |
|---|---|---|---|---|---|---|---|---|---|
| N1 | 80 | 4.6 | 0.1 | 12 | 96 | $2\times 10^6$ | 1.10 | $5.92\times 10^{10}$ | 0.47 |
| N2 | 22 | 3.6 | 0.1 | 48 | 105.6 | $2.8\times 10^6$ | 0.96 | $6.17\times 10^{10}$ | 0.40 |
| N3 | 15 | 2.6 | 0.1 | 66.3 | 99.4 | $1\times 10^6$ | 1.04 | $6.05\times 10^{10}$ | 0.43 |

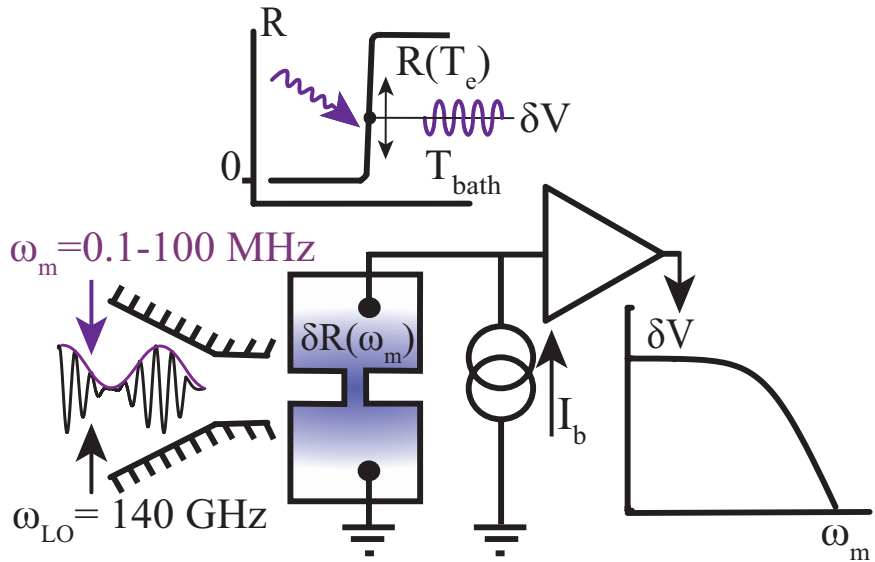

FIG. 1. (a) Schematic block diagram of the experimental setup. The amplitude-modulated radiation from BWO illuminates the sample that is held at temperature in the middle of superconducting transition. An increase of the electron temperature $T_e$ caused by the radiation leads to a proportional increase of the resistance of sample that produces a voltage signal proportional to the bias current $\delta V = I \times \delta R$.

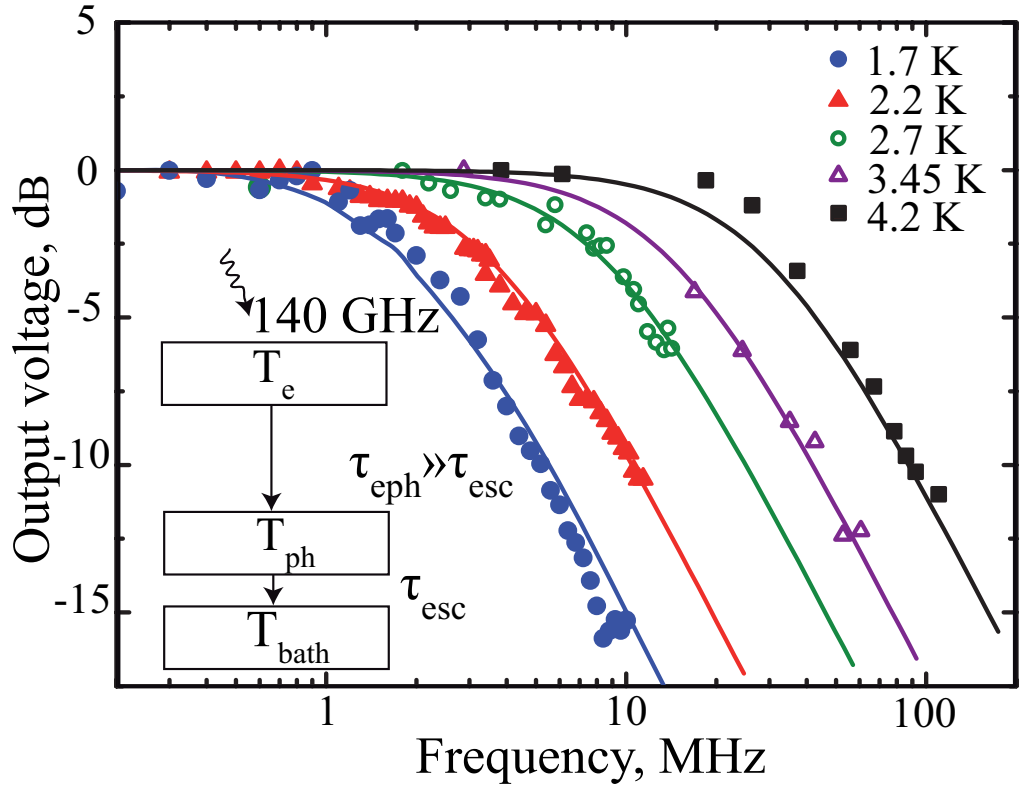

FIG. 2. The frequency dependence of the sample N1 response at different temperatures. The data for each curve was normalized to 0 dB for convenience. The solid lines are the least-squares fit to Eq.3. The fit standard error of the roll-off frequency is not exceeding 10 %. The inset shows the diagram of energy relaxation in sample.

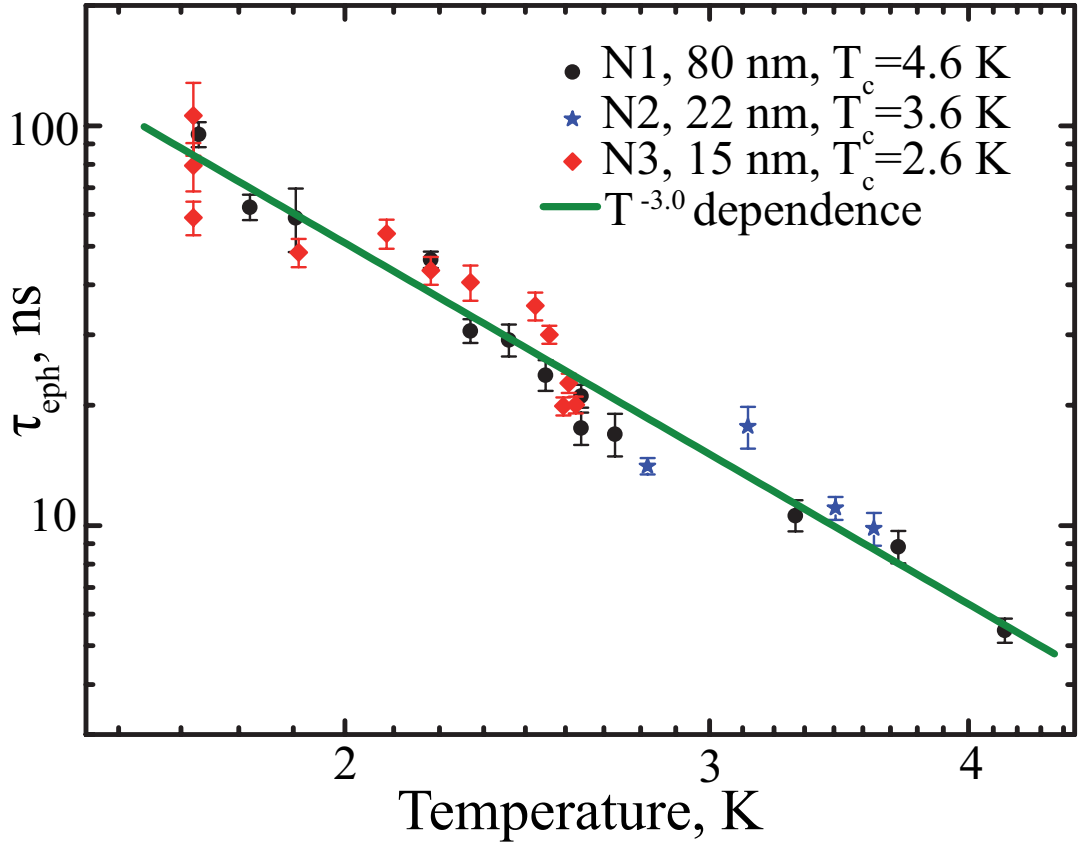

FIG. 3. The electron energy relaxation time for TiN samples with thicknesses 15 nm (diamonds), 22 nm (stars) and 80 nm (circles). The solid line is the least-squares fit that corresponds to $\tau_{eph} = \alpha T^{-n}$, where $n = 3.0 \pm 0.13$ and $\alpha = 407\ ns \times K^3$.